\documentclass[aps,pra,twocolumn,groupedaddress,showpacs]{revtex4}

\newcommand{\eq}[1]{Eq.~(\ref{eq.#1})} 

\newcommand{\fig}[1]{Fig.~\ref{fig.#1}}

\newcommand{\sect}[1]{Sec.~\ref{sect.#1}}

\newcommand{\sectlabel}[1]{\label{sect.#1}}
\newcommand{\eqlabel}[1]{\label{eq.#1}}
\newcommand{\figlabel}[1]{\label{fig.#1}}

\newcommand{\expect}[1]{ \left< #1 \right> }

\newcommand{\ket}[1]{ {\left| #1 \right\rangle} }

\newcommand{\utility}{Q}

\newcommand{\Jpair}{ {J_{\rm pair}} } 

\begin{document}

\title{A Practical Quantum Mechanism for the Public Goods Game}

\author{Kay-Yut Chen}
\author{Tad Hogg}
\author{Raymond Beausoleil}
\affiliation{HP Labs
    \\Palo Alto, CA 94304}

\date{\today}

\begin{abstract}
Quantum generalizations of conventional games broaden the range of
available strategies, which can help improve outcomes for the
participants. With many players, such quantum games can involve
entanglement among many states which is difficult to implement,
especially if the states must be communicated over some distance. This
paper describes a quantum mechanism for the economically significant
$n$-player public goods game that requires only two-particle
entanglement and is thus much easier to implement than more general
quantum mechanisms. In spite of the large temptation to free ride on
the efforts of others in this game, two-particle entanglement is
sufficient to give near optimal expected payoff when players use a
simple mixed strategy for which no player can benefit by making
different choices. This mechanism can also address some heterogeneous
preferences among the players.
\end{abstract}


\maketitle

\section{Introduction}

Quantum information processing provides a variety of new capabilities
with potentially significant performance improvements over
conventional techniques. One example is quantum computation with its
ability to rapidly solve problems, such as factoring~\cite{shor94},
which appear to be otherwise intractable.  However, implementing
machines with enough bits and coherence time to solve problems
difficult enough to be of practical interest is a major
challenge. Another application, quantum cryptography, is feasible
today for exchanging keys over distances of tens of kilometers. A
third area, which potentially can come into practical use soon, is
quantum economic mechanisms and games.  Extending classical games into
the quantum realm broadens the range of strategies~\cite{meyer99}, and
has been examined in the context of the Prisoner's
dilemma~\cite{eisert98,eisert00,du01,du02} and the $n$-player minority
game~\cite{benjamin01}. Quantum games do not require long sequences of
coherent operations and hence are likely to be easier to realize than
large-scale quantum computations.

In this paper, we present a quantum version of an important social
dilemma: public goods. Provisioning for public goods is a well-studied
social choice problem. A typical example is a group deciding whether
to provide a common good, such as a park, in the face of potential
free riders. The free rider problem~\cite{Hardin68} cannot be solved
with traditional means without either a third party to enforce
agreements or a repeated game scenario in which participants can
self-police. Government is one typical solution. While government is a
good solution to public goods involving a large population such as
national defense, it is inefficient to provide public goods in smaller
groups such as neighborhood watch. Provision of these smaller scale
public goods often relies on altruism and other weaker
incentives. Invariably, contributions to these public goods are not at
efficient levels.

Quantum mechanics offers the ability to solve the free-rider
problem in the absence of a third party enforcer in a single shot
game without repetition. With suitable design, simple mixed
strategies almost entirely avoid the free rider problem and give
expected performance close to the Pareto efficient value when the
size of the group is large. In our case, the power of the quantum
mechanism comes from entanglement. Quantum entanglement allows
individuals to pre-commit to agreements where otherwise it would
be individually rational to renege.

Three different quantum mechanisms with different degrees of
entanglement are reported in this paper. These differ in their
performance characteristics and ease of implementation. The
results provide information about how one can best design a
quantum mechanism to solve the free-rider problem.

Equal to the importance of its economics properties is whether a
quantum mechanism can be built. This issue is addressed by
restricting our attention to quantum systems that can be
practically implemented by technologies that could soon be
available. Creating and communicating highly-entangled states
among $n$ players poses significant implementation challenges. The
most readily implementable mechanisms are those that only require
entanglement among pairs of states. Thus an interesting practical
question for developing applications of quantum information
processing is whether performance of the public goods game can be
significantly improved with quantum mechanisms limited to using
two-particle entangled states. We have developed a mechanism only
requiring communication of two-particle entangled states among the
players. This mechanism could be feasible to implement in the near
future even for players at different locations.

The paper is organized as follows. Section 2 describes the general
approach to ``quantize'' a classical game. Section 3 discusses the
economics of the public goods game. Section 4 describes the
quantum version of the public goods game and its solutions.
Section 5 concludes with possible extensions to our mechanism.

\section{Quantum Games}

This section describes one approach to generalize conventional
games to make use of entangled quantum states. We then discuss
issues involved in their implementation, particularly the
significant benefit for games restricted to use only two-particle
entanglement.

\subsection{Creating Quantum Games}

A game consists of a set $S$ of choices available to the players
and an associated payoff to each player based on those choices.
With $s_k \in S$ denoting the choice made by player $k$ and
$s=(s_1,\ldots,s_n)$. A game is defined by the payoffs to the
players depending on these choices, i.e., $P_k(s)$.

One approach~\cite{eisert98,benjamin01} for generalizing these
games to quantum operators considers superpositions of all
possible choices $\sum_{s} \psi_s \ket{s}$ summing over all
choices in $S$ for each player. The quantum version of the game
starts by creating an initial superposition. Subsequently each
player is allowed to operate only on their corresponding part of
the state. In the last stage, the final superposition is used to
produce a definite choice for each player via a further joint
operation followed by a measurement. The initial and final
operations, acting on the full superposition, are fixed and known
to the players.

To give a direct generalization of the original game, the player's
operations should include choices that correspond to the original
choices. That is, the initial and final operations on the state
should reproduce the payoff structure for the original game if all
players restrict their individual operations to just those
corresponding to the actions allowed in the original game.

More precisely, the game proceeds as follows:
\begin{itemize}
\item Starting with a particular initial superposition $v$, create the entangled state
$J v$, where $J$ is an entanglement operator that commutes with
the classical single-player operators.

\item Players select an operation to apply to their part of the superposition, giving $ v' = (U_1 \otimes \ldots \otimes U_n) J v $
where $U_k$ is operator used by player $k$.

\item Finally undo the initial entanglement, giving $\psi =
J^\dagger v'$. For a given game, i.e., choices for $v$ and $J$,
the final superposition is a function of the players' choices,
i.e., $\psi(U_1,\ldots,U_n)$.

\item Measure the state, giving a specific value for each player's choice. The probability to produce choices $s$ (i.e., a
particular assignment, 0 or 1, to each bit) is $|\psi_s|^2$.
\end{itemize}

The choice of $J$ determines the type and amount of entanglement
among the players. The commutation condition on $J$ ensures that
if each player selects the operator corresponding to one of the
choices in the original game, the final result of the quantum game
will, with probability 1, reproduce those choices.

\subsection{Implementing Entanglement for Many Players}

Ideally, we would like our scheme to rely on the distribution of
entangled states between distant players, implying that the qubits are
encoded in the polarization states of photons transmitted throughout a
fiber-optic network. Given a bright source of polarization-entangled
photon pairs~\cite{kwiat99}, these qubits can be delivered by
propagation through optical fibers, and purified using high-quality
linear optical elements~\cite{pan01}. In principle, maximally
entangled $n$-photon states can be constructed from entangled
two-photon states~\cite{bouwmeester99,pan01a}, and these states can be
further manipulated using linear optical elements to perform universal
gate operations~\cite{knill01}.

However, scaling a fully entangled game from 2 to $n$ players can be
nontrivial even when linear optics is used. Suppose a trial between
any two players succeeds with probability $\beta$ (incorporating the
net efficiency with which entanglement can be created, distributed,
purified, manipulated, and detected), so the mean number of trials
needed to successfully register a mutual choice between two players is
$1/\beta$. Because an accidental (or deliberately disruptive)
measurement of a single qubit in the $n$-particle maximally entangled
state destroys the entire state, we expect the number of trials needed
to complete a maximally-entangled game for $n$ players will scale no
better than $\beta^{-n}$. Suppose instead we implement the game by
distributing entangled two-particle states between either all
enumerated pairs of players or nearest neighbors, as described in
\sect{twoway} and \sect{nn}, respectively. In these cases, we expect
that the mean number of trials needed to complete the game will scale
as either $n (n-1)/2 \beta$ or $n/\beta$, and are therefore relatively
easier to implement for games with a large number of players.

For example, in the simplest near-term implementation, a single
game system can be constructed at a central location, and players
can travel to the game and individually specify the operators to
be applied to their qubits. As the technology evolves, the
necessary hardware for specification of qubit operations can be
distributed to distant players, who then can apply their operators
to photonic qubits transported to them over an optical network. In
either case, entangled pairs can be generated and distributed
consecutively until all players have successfully registered a
choice for each pair in which they are a member. Although great
strides continue to be made in multi-particle
experiments~\cite{bouwmeester99,pan01a}, it is clear that ---
until $\beta \longrightarrow 1$ --- two-particle games are far
more feasible, and could allow tests of quantum game theory to be
performed in the near future.

Given some single-trial success probability $\beta$, the number of
trials is limited by the rate at which two-particle entangled
qubits can be provided. A bright source of entangled photon pairs
has been constructed using an argon-ion laser and parametric
down-conversion in BBO crystals, capable of producing 140 detected
two-photon coincidences per second per milliwatt of Ar$^+$ pump
power~\cite{kwiat99}. In principle, given an electrically-driven
source of single photons~\cite{yuan02}, entangled photon pairs
also could be generated in a compact all-solid-state system using
down-conversion in periodically-poled lithium niobate
waveguides~\cite{tanzilli01,mason02}. However, in the future it is
possible that up to $10^9$ pairs per second could be produced
using a single quantum dot embedded in a
\emph{p}-\emph{i}-\emph{n} junction surrounded by a
microcavity~\cite{benson00}.

\section{Public Goods Economics}

\subsection{Overview}

A pure market economy fails to provide efficient levels of public
goods for two key reasons. By definition, a public good is
non-excludable. Once the good is provided, there is no means of
charging for it or restricting access to it. This creates the free
rider problem in which people are tempted to use the public good
without paying for it. The prisoner's dilemma is a perfect
illustration of this free-rider problem. In this two-person game,
each player has the choice to ``cooperate'' and ``defect''.
Payoffs for both players are higher when both of them choose to
cooperate instead of defect. However, each individual is better
off by defecting.

Furthermore, even if there exists a third party (usually the
government) to enforce contribution to the public goods,
individuals have the incentive to hide their preferences on how
much they value the public good. This information asymmetry makes
it difficult to determine the efficiency of public goods
distribution.

Some of these issues have been addressed in economics literature.
For example, if the public good can be provisioned through a
government, there exist mechanisms to reveal preferences of
individuals~\cite{groves77}. Also, experimental work on public
goods~\cite{ledyard95} compares people's actual behaviors to the
predictions of game theory.

The free rider problem, however, is more difficult to overcome. In the
absence of a benevolent dictator, self-motivation becomes the dominant
factor. Luckily, two phenomena mitigate the effects of free
riding. The first is the folk theorem~\cite{rasmusen89,aumann92}. If
the game is played repeatedly within a relatively small group, the
folk theorem suggests an efficient outcome may be enforceable through
the strategy of punishing a defector.  It is arguable whether this
will work in practice because game theory rationality places a strong
burden on the individuals to determine the correct strategies. The
second phenomenon is individuals' motivations may not be completely
selfish: experimental evidence suggests people may be altruistic, at
least in relatively small groups~\cite{ledyard95}. However, ensuring
an efficient outcome is not possible without the intervention of a
third party.

\subsection{Formulating Public Goods Games}

For simplicity in discussing the public goods game, we assume there is
only one public good and one private good which players can use to
contribute to producing the public good. It is easy to generalize to
multiple goods.

There are $n$ players indexed by $k$. We make the following
definitions:
\begin{description}
\item[$x$]      amount of public good
\item[$y_k$]        initial endowment of private good of player $k$
\item[$c_k$]        contribution of player $k$
\item[$\utility_k(x,y)$]   utility of player $k$ when consuming $x$ units of public good and $y$ units
        of private good
\item[$g(C)$]   production function of the public good as a function of total contributions $C = \sum_k c_k$
\end{description}

If contribution is voluntary and continuous, each individual would want to choose a contribution to maximize:
\begin{equation}
\max_{c_k} \utility_k(g(c), y_k-c_k)
\end{equation}
which leads to
\begin{equation}
\frac{1}{d g/d c} = \frac{d\utility_k/dx}{d\utility_k/dy}
\end{equation}
for all $k$ when evaluated at the maximizing choices with $x=g(c)$
and $y=y_k-c_k$. These give $n$ equations for the $n$ contribution values $\{c_k\}$.

This condition says each person will contribute up to the point
where the marginal rate of substitution is equal to the marginal
benefit of his contribution in providing the public good.

We use the standard economic efficiency measure of Pareto
optimality~\cite{kreps90}. That is, there exists no other allocation
such that one player is strictly better off while all others are at
least as well off as before. In our context, Pareto efficiency
requires~\cite{kreps90}
\begin{equation}
\frac{1}{d g/d c} = \sum_k \frac{dU_k/dx}{dU_k/dy}
\end{equation}

Let $C$  be the Pareto efficient level of total contribution and $C'$
be the equilibrium level of total contribution. The above two conditions mean that $g'(C)<g'(C')$.

For well-behaved $g$ with diminishing rate of return, $C>C'$.  Thus
the equilibrium level of contribution is less than the efficient
level.  This analysis assumes the players' contribution choices are
continuous.  However, similar results follow if contributions are
restricted to discrete levels.  For less well behaved $g$, there may
be multiple equilibria as well as contribution levels at the efficient
levels.

\subsection{An Example }

We use a simple example that illustrates the core issue of the
free-rider problem. Assume $\utility_k(x,y) = x + y$ for all $k$
and $g(c) = aC/n$ where $a$ is a parameter and $C$ is the total
contribution level. That is, $C=\sum_k c_k$.

It can be shown quite easily that the following characterizes the
unique Nash equilibrium:
\begin{itemize}
\item If $a<1$, $C=0$ and this is the Pareto efficient outcome

\item If $1<a<n$, $C=0$, but is an inefficient outcome.
One efficient outcome in this case is $c_k=y_k$ but this is not an
equilibrium since each player increases payoff by defecting, i.e.,
switching to $c_k=0$.

\item If $n<a$, $c_k=y_k$ is the efficient outcome.
\end{itemize}

This analysis can be interpreted as follows. The production
function $g$ multiplies the total contribution by $a$. The result
is then equally divided back to the players. If $a$ is less than
1, there is no gain to produce the public good and so the
efficient outcome is {\em not} to produce any. If $a$ is greater
than $n$, then for each unit the player receives back more than
the contribution, thus it is advantageous to contribute and the
equilibrium will be efficient.

The interesting case, giving a social dilemma, is when $a$ is between
1 and $n$. In this case, the public good per person increases with
contribution. However, the marginal benefit of each contribution is
still smaller than 1. Thus a player receives only $a/n$ in benefit for
a unit of additional contribution, which is a net loss. Therefore, it
is rational {\em not} to contribute.  However, failure to contribute
is an inefficient outcome. Thus we have a social dilemma in that the
group as a whole is better off if all contribute, but each person
prefers not to contribute and hence their rational choices lead to no
public good production.  Moreover, this case has multiple Pareto
efficient outcomes. For example, both total contribution and total
contribution from all but one person are efficient outcomes.

\subsection{Heterogeneity and Asymmetric Information\sectlabel{heterogeneous}}

In the case of total contribution, it is also
easy to show that for some set of $\{y_k\}$, one or more
individuals will be worse off than the case of no contribution.

Thus an efficient outcome may not be desirable for other reasons,
such as voluntary participation constraints (some players do not
want to play the game). To address this issue, we will focus our
attention on a smaller set of efficient outcomes that also satisfy
the voluntary participation constraints. Thus, in additional to
Pareto efficiency, we also require $\utility_k(g(C),y_k-c_k) \geq
\utility_k(g(0),y_k)$ for all $k$, i.e., each person will also be
better off in this efficient outcome then the no contribution
case. For our example, this implies
\begin{equation}\eqlabel{voluntary}
\frac{a}{n} \sum_{j=1}^{n} c_j \geq c_k
\end{equation}
for all $k$. The following contribution profile is efficient and
satisfies \eq{voluntary}:
\begin{equation}
c_k=\cases{y_k & if $y_k<C^*$ \cr C^* & if $y_k \geq C^*$ }
\end{equation}
where
\begin{equation}
C^* = \frac{a}{n - a n + a m} \sum_{j=1}^{m} y_j
\end{equation}
where $y_k$ is sorted in ascending order and $m$ is the largest
integer less than $n$ for which $C^* \geq y_k$ holds for all
$k=1\ldots m$.

Under this additional constraint, if the distribution of wealth is
narrow (specifically, $a \bar{y} \geq y_k$ for all $k$ where
$\bar{y}=\frac{1}{n} \sum_k y_k$ is the mean value of the private
goods), then everyone should contribute everything. If there is a
wider distribution of wealth, then there is a cut-off point $C^*$.
Everyone should contribute everything if their wealth $y_k \leq
C^*$ and contribute only up to $C^*$ if their wealth is more than
$C^*$. Thus to maintain voluntary participation the rich should
contribute more in absolute terms than the poor, but less in
percentage terms.

If wealth is distributed narrowly, (satisfying \eq{voluntary} if
individual contributes everything) then there is no need for
asymmetric contribution. Therefore, it is sufficient to treat the
problem as if wealth is equal.

However, if condition \eq{voluntary} is not satisfied, a new
incentive issue arises. To be able to solicit the ``correct''
amount of contribution from every individual, we not only need to
solve the free-rider problem, but also correctly identify the
wealth level of every individual. Furthermore, individuals have
incentives to pretend to be poorer than they are to minimize their
contributions.

The following example illustrates these issues. Consider a
population with two levels of wealth: $m$ individuals have initial
wealth $y$ and $n-m$ individuals have wealth $\alpha y$ where
$\alpha > 1$. We are interested in the issue of whether a
mechanism can achieve an equilibrium that is not only efficient,
but also satisfies the voluntary participation constraint. The
only interesting cases are where the contribution strategy is
asymmetric. That is, not every individual has to contribute
everything in the desirable allocation. One interesting case is
where $m=n-1$ (with only {\em one} high-wealth individual). In
this case, it can be shown that the desirable allocation is
everyone with the lower wealth contributes everything. The person
with wealth $\alpha y$ should not contribute everything if $\alpha
> \frac{a(n-1)}{(n-a)}$. He should contribute
$\frac{ay(n-1)}{n-a}$.

\section{A Quantum Mechanism for Public Goods Provisioning}

We first characterize equilibria of the quantum game of the
homogeneous version of the public goods game. This allows us to
study several configurations, such as different entanglement and
interpretation of the qubits, of the quantum game. Subsequently,
the results in the simple homogeneous case will be extended to the
heterogeneous case.

For the quantum mechanism, each player can choose either to
contribute nothing ($c_k=0$, ``defect'') or everything ($c_k=y$,
``cooperate''). We can also consider an intermediate case in which
players can select from a discrete range of contribution values,
$0, y/K, 2y/K, \ldots, y$ for various choices of $K$, but in our
case allowing such intermediate contributions gives lower average
payoffs for the strategies we present below.

Here is an example of the intermediate case. For $n=3$ players and
using 3 bits to specify discrete choices: either contribute fully
($c_k=y$, ``cooperate'') or contribute nothing ($c_k=0$,
``defect''), there are 8 states. Suppose we let the value 0
correspond to ``cooperate''. Then the payoffs to the three players
are (using $y=1$)

\begin{equation}
\begin{array}{cccc}
000 & a         & a         & a\\
001 & 2a/3      & 2a/3      & 2a/3+1 \\
010 & 2a/3      & 2a/3+1    & 2a/3 \\
011 & a/3       & a/3+1     & a/3+1 \\
100 & 2a/3+1    & 2a/3      & 2a/3 \\
101 & a/3+1     & a/3       & a/3+1 \\
110 & a/3+1     & a/3+1     & a/3   \\
111 & 1         & 1         &   1
\end{array}
\end{equation}

The quantum version of the game is set up as follows: first create
entangled qubits (with 0 and 1 representing cooperate and defect,
respectively), allow the individuals to operate on their
individual qubit, then combine the result (by undoing the initial
entangling operation). To preserve the correspondence with the
original game, the entanglement operator should commute with those
quantum operations corresponding to the classical choices.
The final measurement gives a definite value for each qubit, which
then corresponds to the individuals' choices.

In general, players are allowed to apply any operator to their
qubit(s). We consider general single-qubit operators, given by
\begin{equation}\eqlabel{U}
U(\theta, \phi, \alpha) = \pmatrix{
 e^{-i \phi} \cos \frac{\theta}{2}   &   e^{i \alpha} \sin \frac{\theta}{2} \cr
 -e^{-i \alpha} \sin \frac{\theta}{2}  &  e^{i \phi} \cos \frac{\theta}{2} \cr
}
\end{equation}
up to an irrelevant overall phase factor. (A further
generalization would allow measurements on the single qubit. This
gives no advantage in at least in some cases~\cite{benjamin01}.)

For $n=2$, this reduces to the Prisoner's dilemma, which has a nice
interpretation in terms of conventional mechanisms. Entangled states
allow player 1 to affect the final outcome produced by the action of
player 2 and vice versa. In a way, it allows for
pre-commitment. Consider the following argument. Player 1 would love
to tell player 2 that if player 2 commits to cooperate, then he would
also cooperate. However, without playing a repeated version of the
game, the ability to punish the other player or without a 3rd party to
enforce the commitment, both players will realize immediately they are
better off reneging their commitments. Entanglement allows the parties
to commit without a third party to enforce the commitments.

The expected payoffs can be viewed as functions of the players'
choices and game definition: $P_k(U_1,\ldots,U_n; J, a)$ (where we
take $y=1$ without loss of generality since it just rescales the
payoffs).

\subsection{Equilibria for the Quantum Public Goods Game}

In this section, we characterize equilibria for three schemes of
entanglement of the public goods quantum mechanism. If players are
allowed to use any single qubit operators given by \eq{U}, there
is no single pure strategy equilibrium. However, we found  mixed
strategy equilibria, with expected payoffs depending on the degree
of entanglement provided in the initial state. In each case, we
find multiple equilibria. These payoffs are superior to that
produced by the classical game in which all players defect so no
public good is produced.

We assume all individuals are risk-neutral expected utility
maximizers. Player $k$'s expected payoff function is given by
$P_k(\psi) = \sum_s P_k(s) |\psi(s)|^2$ where $P_k(s)$ is the
payoff for player $k$ given the choices specified by state $s$.

We use the Bayesian Nash equilibrium as the solution concept for
the quantum game. Each individual will play a strategy (pure or
mixed) such that they are mutually maximizing their expected
payoff. None has the incentive to make a unilateral change to
their strategy.

A single-player operator $\hat{u}$ forms a symmetric Nash
equilibrium if for any other choice $u \neq \hat{u}$
\begin{equation}
P_k(\psi(\hat{u},\ldots,\hat{u})) \geq
P_k(\psi(\hat{u},\ldots,\hat{u},u,\hat{u},\ldots,\hat{u}))
\end{equation}
for all players $k$, with $u$ substituted for the $k^{th}$
player's choice on the right-hand side. For homogeneous
preferences, it is sufficient that this hold for just one player.
More generally, asymmetric equilibria involve possibly different
operations for each player.

Whether such an equilibrium exists, and if so whether it is unique
and gives the optimum payoffs for the players, depends on the set
of allowed operations, the amount and type of entanglement
(specified by the choice of $J$) and the nature of the payoffs.

Our analysis includes mixed strategy equilibria since in many
cases, particularly with respect to the quantum version of the
public goods game, there is no pure strategy equilibria. The
strategic space for quantum games are infinite. We limit our
attention to finite mixed strategies. That is, we only allow
individuals to randomly (with any probabilities assignment) choose
within a finite set of operators. We also make the standard
assumption that individuals have access to a perfect randomization
process.

In the next three subsections, we report three different schemes of
entanglement and their corresponding mixed-strategy Nash
equilibrium.

\subsection{Full Entanglement}

A conceptually simple approach allows arbitrary entanglement among the
players' qubits. As one example, consider fully entangled states. The
initial entangled state is $(\ket{00...0}+i \ket{11...1})/\sqrt{2}$,
using the $2^n \times 2^n$ entanglement matrix
\begin{equation}\eqlabel{J}
J_n = \frac{1}{\sqrt{2}} (I + i \sigma_x \otimes \ldots \otimes \sigma_x)
\end{equation}
where the product in the second term consists of $n$ factors of
$\sigma_x$, the $2 \times 2$ Pauli matrix $\pmatrix{0 & 1 \cr 1 & 0
}$.

Allowing general single-bit operators of \eq{U}, we find no pure
strategy Nash equilibrium for the players. However, there are a
variety of mixed strategy equilibria. As one example, let
\begin{eqnarray}\eqlabel{mixture}
u(0) &\equiv&   U(0,0,0) = \pmatrix{1 & 0 \cr 0 & 1 \cr} \\
u(1) &\equiv&   U(0,\pi/2,0) = \pmatrix{i & 0 \cr 0 & -i \cr} \nonumber
\end{eqnarray}
Note $u(0)$ corresponds to the classical ``cooperate'' option.  A
mixed strategy consisting of each player randomly selecting $u(0)$ or
$u(1)$, each with probability $1/2$, gives expected payoff of
$(1+a)/2$. This is an equilibrium: if any one player switches to using
a different operator, or different mixture of operators, the expected
payoff for that player remains equal to $(1+a)/2$.  While this payoff
is less than the efficient outcome, it is substantially better than
the classical outcome with payoff of 1 since all choose to defect.

Although this scheme is not practical with respect to implementation
due to its use of highly entangled states, we include it as a
comparison to other schemes.

\subsection{Two-particle Entanglement\sectlabel{twoway}}

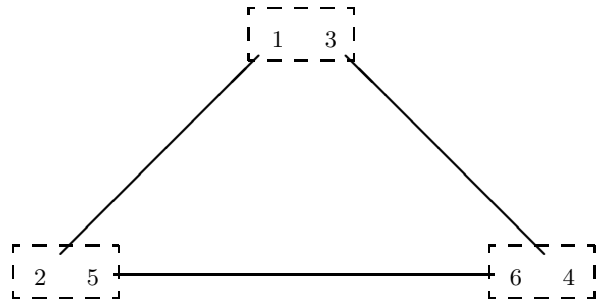
\begin{figure}
\begin{center}
\begin{picture}(200,100)(-100,0)
\thicklines

\put(-10,90){1} \put(10,90){3}

\put(-100,0){2} \put(-80,0){5}

\put(100,0){4} \put(80,0){6}

\put(-70,4){\line(1,0){144}} \put(-90,12){\line(1,1){75}}
\put(93,12){\line(-1,1){75}}

\thinlines \put(-19,85){\dashbox{5}(40,20){}}
\put(-108,-5){\dashbox{5}(40,20){}}
\put(72,-5){\dashbox{5}(40,20){}}
\end{picture}
\end{center}
\caption{\figlabel{pairs}Six qubits giving two-particle
entanglement among three players. The first player operates on
bits 1 and 3, which are entangled with bits 2 and 4, respectively
owned by the second and third players.}
\end{figure}

Full entanglement is difficult to implement as $n$ increases,
particularly for qubits communicated over long distances. Thus we
consider restricting entanglement to only pairs of qubits. In
this case, we suppose each pair of players has a maximally
entangled pair, so each player has $n-1$ qubits.

The entanglement matrix for a case consisting of $N={n \choose 2}$ pairs is
\begin{equation}\eqlabel{Jpair}
\Jpair(N) = J_2 \otimes \ldots \otimes J_2
\end{equation}
with the product consisting of $N$ factors of the entanglement
operator of \eq{J} for the case of $n=2$, i.e., full entanglement
among two qubits.

With multiple bits per player, we also need to specify how the final
measured state is to be interpreted. One approach is to allow various
amounts of contribution rather than all or none. That is, if $z$ of
the $n-1$ bits for player $k$ are 0, player $k$'s contribution is $c_k
= y z/(n-1)$, ranging from 0 to $y$. So instead of two choices, the
player has a range of possible contributions. This choice gives the
same result as the fully entangled case: the mixed strategies have
expected payoff of $(1+a)/2$ and remain weak equilibria.

For example, $n=3$ uses six qubits corresponding to the pairs of
players $(1,2)$, $(1,3)$ and $(2,3)$, as shown in \fig{pairs}.
Thus, for example, the first player operates on the first and
third qubit in this ordering of the bits. The state
$\ket{0,0,0,1,1,1}$ has $0,0$ for the first and third qubit, so
the first player has $z=2$ and contributes $y$. The second player,
using the second and fifth bits, has $0,1$ with $z=1$ and
contributes $y/2$.

An alternate interpretation of the bits provides higher payoffs while
maintaining the same mixed strategy equilibria. Specifically, we again
suppose contributions are all or nothing but now consider the player
to contribute if {\em any} of the $n-1$ bits equals 0. This simple
change in the construction of the game gives expected payoff equal to
\begin{equation}\eqlabel{payoff}
a - 2^{-(n-1)}(a-1)
\end{equation}
which, since $1<a<n$, is only slightly less than
the highest possible payoff, $a$. As examples, the expected payoffs for $n=3$
and $4$ are, respectively, $(1+3a)/4$ and $(1+7a)/8$. As $n$
increases, the expected payoff approaches the optimal value.

We could also consider other interpretations, e.g., full contribution
if a majority of the bits are 0, and otherwise no contribution.

Significantly, the mixed strategy remains an equilibrium even if a
player applies different operators to each of the $n-1$ bits.

\subsection{Two-particle Entanglement with Neighbors\sectlabel{nn}}

Two-particle entanglement among all possible pairs of players
requires $n(n-1)/2$ entangled pairs. While significantly easier to
implement than entanglement among $n$-players, we can also
consider behavior with even less entanglement. Specifically,
consider the players in some arbitrary order and only provide an
entangled pair between successive players in that order (with an
additional pair between the first and last). This entanglement
requires only $2n$ qubits.

This case maintains the same equilibrium mixed strategies. If we
interpret the two bits of each player as allowing partial
contributions, the expected payoff remains $(1+a)/2$. Using the
all-or-none method, where a player contributes everything if at
least one of the two bits equals 0, the payoff is $(1+3a)/4$ for
all $n$. Note this is the same as the payoff of the full
two-particle case, \eq{payoff}, for $n=3$ (as expected: for $n=3$
the neighbor pairs are the same as a two-particle entanglement
between all pairs of players).

Again, the payoff is superior to the classical game Nash
equilibrium. Unlike entanglement among all pairs, the payoff does not
improve with larger $n$. Thus this result illustrates a tradeoff:
lower performance when using fewer pairs.

\subsection{Generalization to Heterogenous Individuals}

The issue of heterogenous wealth is largely ignored in our analysis of
the quantum game. If the distribution of wealth is narrow (as defined
in \sect{heterogeneous}), an efficient quantum solution that assumes
homogeneous wealth will also satisfy the voluntary participation
constraints making heterogeneity a non-issue.

Furthermore, if the issue of adverse selection (incentive to hide
information) is addressed by some other method, then the quantum
mechanism can be used in tandem to address the general case.
Specifically, in the case of heterogenous wealth,
if every individual's wealth is revealed to the mechanism, then
the mechanism can be modified slightly by the following method to
yield the desirable outcome. First calculate the optimal
contribution for every individual based on the revealed wealth
levels as described in \sect{heterogeneous}. Then the players play the
quantum game with the knowledge that the final qubits are
interpreted as follows: an individual contributes the optimal
amount, not his total wealth, if one or more of his qubits are
zero. Essentially, all the contribution levels are pre-determined
and the issues reduce to just the free-rider problem.

\section{Conclusion}

Quantum mechanics can be used to develop new formulation of
classical economics games which give arise to new solutions. In
this paper, we have shown how a quantum mechanism can be
constructed to solve the free-rider problem in the public goods
game, without the need of third party enforcement nor repeated
play. Implementation issues are also explored and addressed.

Most of the power of this new mechanism comes from entangled
states, which in theory allow individuals to co-ordinate and
commit in environments when classical means do not. Incidentally,
entanglement is also the major issue determining whether a quantum
mechanism is practical or not.

Three different schemes of entanglement are explored. We found
that two-particle entanglement, which is feasible for the near
future, can also solve the free-rider problem and achieve nearly
efficient outcomes. Furthermore, we have also argued that the
mechanism is robust with respect to limited amount of
heterogeneity in the system if there is no adverse selection.

Game theoretic solutions (such as the Bayesian Nash equilibrium we
discuss in this paper) are at best approximations of real human
behavior. In this case, rationality dictates that each individual has
a full understanding of the quantum mechanical implications of his
choices. How well this describes the actual behavior of people
involved in quantum games is an interesting direction for future
work with laborabory experiments involving human subjects.

There are many natural extensions of this research.  First, people
may use criteria other than expected payoff, e.g., to minimize
variance in payoff if they are risk adverse.  Second, the case of
heterogeneous players and adverse selection requires further
analysis.  This work also suggests experimental research,
exploring the issues of practicality of implementation and human
behavior with respect to manipulating quantum states.

\section*{Acknowledgments}
We thank Bernardo Huberman, Alan Karp, Phil Kuekes and Wim van Dam
for helpful discussions.

\appendix
\section{Derivation of Mixed Strategy Payoff}

{ 
\newcommand{\Uop}[1]{ u^{(#1)} } 
\newcommand{\vPair}[1]{ v^{(#1)} } 
\newcommand{\psiPair}[2]{ {\psi_{\rm pair}(#1,#2)} }
\newcommand{\Prob}{ {\rm Pr} } 

This appendix derives \eq{payoff} and shows the mixed strategy is
indeed an equilibrium: no single player can benefit from deviating
from the mixture. A similar derivation applies to the other cases with
different entanglement (i.e., full or two-particle only among
neighbors) and interpreting multiple bits per player as indicating
partial contributions. For simplicity, we take the private good value
to be $y=1$.

Consider the behavior of player 1, selecting operators
$\Uop{2},\ldots,\Uop{n}$ while all other players select either $u(0)$ or
$u(1)$ of \eq{mixture} with equal probability for all their bits.
The initial state $\Jpair(N) (1,0,\ldots,0)$ is
\begin{displaymath}
\bigotimes \frac{1}{\sqrt{2}} (1,0,0,i)
\end{displaymath}
with one factor for each pair. Subsequent operations on each pair
are independent. Consider the pair between players $j,k$. If these
players use operators $A$ and $B$ respectively, the final state
for their pair is
\begin{equation}\eqlabel{pair}
\psiPair{A}{B} = J_2^\dagger (A \otimes B) \frac{1}{\sqrt{2}} (1,0,0,i)
\end{equation}

Players other than the first use either $u(0)$ or $u(1)$. Evaluating
the products in \eq{pair} for these cases gives
\begin{eqnarray}\eqlabel{psiPair}
\psiPair{u(0)}{u(0)}    &=& (1,0,0,0) \\
\psiPair{u(0)}{u(1)}    &=& (0,0,0,1)   \nonumber \\
\psiPair{u(1)}{u(0)}    &=& (0,0,0,1)   \nonumber \\
\psiPair{u(1)}{u(1)}    &=& (-1,0,0,0)  \nonumber
\end{eqnarray}
so players making the same choice produce a pair equal to $\pm
\ket{00}$ (i.e., both cooperate), while those making opposite choices
give $\ket{11}$ (i.e., both defect).

For a given instance of this mixed strategy, let $t_k \in \{0,1\}$
indicate the operator choice of player $k=2,\ldots,n$: $u(t_k)$.
Then the final state for the pair involving players $j$ and $k$ is
just $\psiPair{u(t_j)}{u(t_k)}$. Thus the portion of the final
state corresponding to pairs not involving player 1 is
\begin{displaymath}
\psi_{\rm other} = \bigotimes_{j,k} \psiPair{u(t_j)}{u(t_k)}
\end{displaymath}
with the tensor product over all pairs $2 \leq j < k \leq n$. The
nonzero components of this vector are all $\pm 1$, which have unit
magnitude so do not affect the probabilities of the final
measurement.

The pair involving players 1 and $k$ gives $\vPair{k}(t_k) =
\psiPair{\Uop{k}}{u(t_k)}$. For any choice of operator $\Uop{k}$,
evaluating \eq{pair} using \eq{mixture} gives
\begin{eqnarray}\eqlabel{vPair}
\vPair{k}_{0,0}(1) &=& -\vPair{k}_{1,1}(0) \\
\vPair{k}_{0,1}(1) &=& \vPair{k}_{1,0}(0) \nonumber \\
\vPair{k}_{1,0}(1) &=& -\vPair{k}_{0,1}(0) \nonumber \\
\vPair{k}_{1,1}(1) &=& \vPair{k}_{0,0}(0) \nonumber
\end{eqnarray}
so, apart from some sign changes, player $k$ switching from $u(0)$
to $u(1)$ simply reverses the result of the two-particle
interaction between players 1 and $k$.

The overall final state is the tensor product of these results for
the individual pairs. The nonzero components of this state vector
are specified by the values for the bits involving player 1. That
is, the final state has the form
\begin{displaymath}
\bigotimes_{k=2}^n \left(
     \sum_{x_k,y_k} \vPair{k}_{x_k,y_k}(t_k) \ket{x_k,y_k}
\right)
\otimes \psi_{\rm other}
\end{displaymath}
with the $x_k, y_k$ each summed over 0 and 1.

Measuring this final state produces a state with definite values
for the $x_k,y_k$, with probability $\Prob(x,y,t)=\prod_{k=2}^n
|\vPair{k}_{x_k,y_k}(t_k)|^2$. For this state, we determine the
payoff to player 1 as follows.

First, the all-or-none interpretation of the bits means player 1
contributes $1$ if any of the $x_k=0$. Defining the indicator
function $\chi(p)$ to equal 1 when the proposition $p$ is true and
0 otherwise, we can write this contribution as $1-\prod_k
\chi(x_k=1)$.

The contribution for player $k>1$ is $1$ if it has a 0 bit in its
pair with player 1 (i.e., $y_k=0$) or at least one player (other
than players 1 or $k$) makes the same choice of operator as player
$k$ (since then \eq{psiPair} shows that pair of players will have
value $\ket{0,0}$ so, in particular, player $k$ will have at least
one of its bits equal to zero). Let $n_b$ be the number of players
$2,\ldots,n$ that select operator $u(b)$, for $b=0,1$. Note $n_b$
is the number of values in $t_2,\ldots,t_n$ equal to $b$, and
$n_0+n_1=n-1$. With these definitions, the contribution of player
$k$ is $\chi(y_k=0 \wedge n_{t_k}=1) + \chi(n_{t_k}>1)$.

Combining the contributions from all players, the payoff
$P_1(x,y,t)$ to player 1 for this measured state, is then
of the form $\frac{a}{n} (1+A) + (1-\frac{a}{n}) B$ with
\begin{eqnarray*}
A &=&
  \sum_k \left( \chi(y_k=0 \wedge n_{t_k}=1) + \chi(n_{t_k}>1)
  \right) \\
B &=&
   \prod_k \chi(x_k=1) \nonumber
\end{eqnarray*}
In this expression, $\sum_k \chi(n_{t_k}>1)$ can be written as
$\sum_{t=0}^1 \chi(n_t>1) \sum_k \chi(t_k=t) = \sum_t n_t
\chi(n_t>1)$.

The expected payoff for player 1 for the given choices of the
other players (as specified by the $t_k$ values) is $\sum_{x,y}
\Prob(x,y,t) P_1(x,y,t)$.

Finally, the mixed strategy used by the other players means each of
the $2^{n-1}$ choices for the values of the $t_k$ is equally likely,
and must be summed over to get the expected payoff of player 1 when
the others use the mixed strategy: $\expect{P_1} =
2^{-(n-1)}\sum_{x,y,t} \Prob(x,y,t) P_1(x,y,t)$.

In the sum over $x,y$, only the factor $|\vPair{k}_{x_k,y_k}(t_k)|^2$
in $\Prob(x,y,t)$ depends on $x_k,y_k$. Thus for terms involving
player $k$, the remaining factors in $\Prob(x,y,t)$ sum to 1 since the
$\vPair{k}(t_k)$ are normalized vectors.

Thus $\expect{P_1}$ is a sum of three terms. The first is
\begin{displaymath}
2^{-(n-1)} \frac{a}{n} \sum_t (1 + n_0 \chi(n_0>1) + n_1 \chi(n_1>1))
\end{displaymath}
or
\begin{displaymath}
\frac{a}{n} (1 + (n-1)(1-2^{2-n}))
\end{displaymath}

The second term is
\begin{displaymath}
2^{-(n-1)} \frac{a}{n} \sum_{k,t,x_k,y_k} |\vPair{k}_{x_k,y_k}(t_k)|^2 \chi(y_k=0 \wedge n_{t_k}=1)
\end{displaymath}
For $n_{t_k}=1$, the only terms contributing to the sum over $t$ are
those for which $t_j \ne t_k$, for all $j \ne k$, i.e., there are just two
cases: $t_k=0$ and the rest are 1, and vice versa. So this term
becomes
\begin{displaymath}
2^{-(n-1)} \frac{a}{n} \sum_k \sum_{t=0}^1 \sum_{x_k} |\vPair{k}_{x_k,0}(t)|^2
\end{displaymath}
The inner two sums give $|\vPair{k}_{0,0}(0)|^2 +
|\vPair{k}_{1,0}(0)|^2 + |\vPair{k}_{0,0}(1)|^2 +
|\vPair{k}_{1,0}(1)|^2$ which, from \eq{vPair}, equals $\sum_{x,y}
|\vPair{k}_{x,y}(0)|^2 = 1$ since the $\vPair{k}$ vectors are
normalized. Thus this term is $2^{-(n-1)} \frac{a}{n}(n-1)$.

Similarly, \eq{vPair} gives the third term equal to $2^{-(n-1)} \left( 1-\frac{a}{n} \right)$.

Combining these results, $\expect{P_1}$ is
$a - 2^{-(n-1)}(a-1)$. This result is independent of the operators
selected by player 1, i.e., the values of the $\vPair{k}$.

Other choices for the mixed strategy operators $u(0), u(1)$ are
possible as well. They need only satisfy \eq{psiPair} (up to an
overall phase factor) and also compensate for any choices made by the
first player via \eq{vPair}, again up to overall phase factors.

} 


\end{document}